\documentclass[a4paper,twoside]{article}

\usepackage{epsfig}
\usepackage{subcaption}
\usepackage{calc}
\usepackage{amssymb}
\usepackage{amstext}
\usepackage{amsmath}
\usepackage{amsthm}
\usepackage{multicol}
\usepackage{pslatex}
\usepackage{apalike}
\usepackage{algorithm2e}
\usepackage[bottom]{footmisc}
\usepackage{SCITEPRESS}     

\usepackage{cite}
\usepackage{graphicx}
\usepackage{multirow}
\usepackage[table,xcdraw]{xcolor}
\usepackage{lipsum,booktabs}
\usepackage{hyperref}
\usepackage{listings}
\usepackage{color}
\usepackage{float}
\usepackage{subcaption}
\graphicspath{{./image/}}

\begin{document}

\title{CPE-Identifier: Automated CPE identification and CVE summaries annotation with Deep Learning and NLP}

\author{\authorname{Wanyu Hu\orcidAuthor{0000-0002-7600-808X}, Vrizlynn L. L. Thing\orcidAuthor{0000-0003-4424-8596}}
\affiliation{ST Engineering, Singapore}
}


\abstract{With the drastic increase in the number of new vulnerabilities in the National Vulnerability Database (NVD) every year, the workload for NVD analysts to associate the Common Platform Enumeration (CPE) with the Common Vulnerabilities and Exposures (CVE) summaries becomes increasingly laborious and slow. The delay causes organisations, which depend on NVD for vulnerability management and security measurement, to be more vulnerable to zero-day attacks. Thus, it is essential to come out with a technique and tool to extract the CPEs in the CVE summaries accurately and quickly. In this work, we propose the CPE-Identifier system, an automated CPE annotating and extracting system, from the CVE summaries. The system can be used as a tool to identify CPE entities from new CVE text inputs. Moreover, we also automate the data generating and labeling processes using deep learning models. Due to the complexity of the CVE texts, new technical terminologies appear frequently. To identify novel words in future CVE texts, we apply Natural Language Processing (NLP) Named Entity Recognition (NER), to identify new technical jargons in the text. Our proposed model achieves an F1 score of 95.48\%, an accuracy score of 99.13\%, a precision of 94.83\%, and a recall of 96.14\%. We show that it outperforms prior works on automated CVE-CPE labeling by more than 9\% on all metrics.}

\onecolumn \maketitle \normalsize \setcounter{footnote}{0} \vfill

\section{\uppercase{Introduction}}
\label{sec:introduction}

As the world advances into the digital era, there is a drastic increase in the number of softwares. These modern softwares depend heavily on open-source libraries and packages. Many organizations spend a lot of resources to monitor the new vulnerabilities in those open source softwares. One of the most well-known open-source vulnerability databases is the National Vulnerability Database (NVD), maintained by the National Institute of Standards and Technology (NIST). 

There are thousands of new vulnerabilities disclosed each year. In 2021, an average of fifty \cite{redscan} new vulnerabilities were disclosed daily, and a total of 20169 \cite{NVD_CVE} new vulnerabilities were published in 2021. The number of new vulnerabilities is increasing rapidly. More than 8000 new vulnerabilities were published in the first quarter of 2022, around a 25 percent \cite{Comparitech} increase from the same period in 2021. Many organizations’ vulnerabilities databases heavily depend on the NVD notifications. Therefore, timely analysis of the new open-source vulnerabilities is crucial for an organization. 

However, according to a survey, once a new vulnerability is publicly disclosed by the Common Vulnerabilities and Exposures (CVE) system and published on the NVD, the metadata used in vulnerability management, such as Common Platform Enumeration (CPE) applicability statements, took 35 days on average\cite{Wareus}. This hinders organizations from getting important vulnerability information in time and leaves the organizations vulnerable to zero-day attacks. 

Therefore, this paper proposed an automated CPE annotating system, called the CPE-Identifier system, which automates the CPE extraction process in the CVE summaries and, in addition to the improved F1, accuracy, precision, and recall scores, it increases the potential time performance improvements in the analysis. This system serves as a tool to identify CPE entities from new CVE text accurately, quickly and conveniently. Named entity recognition (NER) is used in the system to automatically identify CPEs in CVE descriptions. This can be useful for security researchers who want to quickly identify the systems and packages affected by a particular CVE. Moreover, while various past machine learning methods were experimented, they require a large amount of annotated training data, which is not publicly available for every CVE summary since the year 1999. Past papers used various methods to annotate the training data. Bridges et al. \cite{Bridges} used "Database Matching, Heuristic Rules and Relevant Terms Gazetteer" to label the CPEs in the sentence, Wareus E. et al. \cite{Wareus} mapped the words in the sentence to the existing CPEs in the available CPE-list. However, these methods are often costly and not a viable solution. Therefore, we also proposed an idea to automate this data generating and labeling process using deep learning models. Our work outperformed previous methods for automatic CPE labeling in all aspects of performance metrics. Our best model obtains a F1 score of 95.48\%, Accuracy score of 99.13\%, Precision score of 94.83\%, Recall score of 96.14\%.

The structure of this paper is as follow: 
\hyperref[sec:Literature Review]{Section 2} introduces the relevant works while discussing their advantages and challenges. \hyperref[sec:Research Methodology]{Section 3} explores the research methodologies, which include the system design and our proposed five NLP models for different purposes in our system. 
\hyperref[sec:Dataset and Labels]{Section 4} elaborates on all the datasets and data engineering processes, including the automated data annotation, generation, and cleaning. 
\hyperref[sec:Models]{Section 5} explains the three SoTA models: BERT, XLNet and GPT-2.
\hyperref[sec:Implementation]{Section 6} describes the implementation of the research, including the Data pre-processing, models training and design of the Graphical User Interface (GUI). 
\hyperref[sec:Result Analysis]{Section 7} analyses the experiment result. 
\hyperref[sec:Conclusion Future Work]{Section 8} concludes the paper by identifying the best model for the CPE-Identifier system and its usability. We also propose directions for future work.

\section{\uppercase{Literature Review}}
\label{sec:Literature Review}
There are past research leveraging NLP methods for CVE summaries. Most of these works are related to Common Attack Pattern Enumeration and Classification (CAPEC), Common Vulnerability Scoring System (CVSS) score, and Common Weakness Enumeration (CWE). K Kanakogi et al. \cite{Kanakogi} have suggested automatically associating the CVE-ID with the CAPEC-ID using the Term Frequency-Inverse Document Frequency (TF-IDF) technique \cite{TF-ID}. Rostami, S. et al. \cite{Rostami} built 12 Multiple Layer Perceptron (MLP) models to predict and map those missing ATT\&CK tactic categories to those CAPEC without tactic types. 

Saba Janamian  et al. \cite{UC} have introduced the VulnerWatch to predict the eight criteria for calculating the CVSS score. Similarly, Shahid, M. R. et al \cite{Shahid} suggested using multiple BERT classifiers for each metric composing the CVSS vector. While Ohuabunwa, B. C. et al. \cite{Ohuabunwa} proposed Categorical Boosting (CatBoost) to predict the CVSS version 3 scores, Evangelista, J. \cite{Evangelista} compared the predictive performance of Latent Dirichlet Allocation (LDA) and BERT in predicting CVSS version 2 scores. Finally, to convert the metrics from CVSS version 2 to version 3, Nowak, M. et al. \cite{Nowak} compared the three models – Naive Bayes classier (NB), k-nearest Neighbors Algorithm (kNN), and Kernel Support Vector Machine (KSVM). Each model is trained on CVE summaries and predicts the CVSS scores.

When dealing with vulnerabilities, Das, S. S. et al. \cite{Das} maps CVEs to CWEs using BERT model to understand the impact and mitigate the vulnerabilities. Stephan Neuhaus et al. \cite{Neuhaus} experimented with a similar idea using the unsupervised LDA model to classify CVE texts into vulnerabilities. Yang H. et al. \cite{Yang} used multiple ML models to analyze the number of CVE References containing PoC code and how these References affect the CVE exploitability. Similarly, Yin J. et al. \cite{Yin} performed similar research using a Bert-based model with a Pooling layer. Bozorgi M. et al. \cite{Bozorgi} also researched the vulnerability's likelihood and exploitability using the Support vector machines (SVMs) method.

To extract the CPE features from the NVD CVE summaries. Huff P. et al. \cite{Huff} proposed a recommendation system using Random Forest and Fuzzy Matching to automatically match the company's hardware and software inventory in the CVE summary to CPEs' Vendor and Product name. However, the size of the data is insufficient. Jia Y. et al. \cite{Jia} proposed to use multiple datasets and the Stanford NER to extract cybersecurity-related entities and train an extractor to extract cybersecurity-related CPE entities and build a knowledge base. However, the F1 score is only below 80\%. Georgescu et al. \cite{Georgescu} uses IBM Watson Knowledge Studio services to train the CVE IoT dataset to enhance the diagnosing and detect possible vulnerabilities within IoT systems. To extract vulnerability types from CVE summaries, Bridges et al. \cite{Bridges} proposed a summarization tool called CVErizer to summarise CVE texts and extract vulnerability types. The author used Auto-Labelling rules to link CVE word with CPE entities through CWE classification. However, the Auto-Labelling rules, like the Heuristic Rules, cannot provide an accurate result. Lastly, Wareus E. et al. \cite{Wareus} have suggested using the Bi-directional Long Short-Term Memory (Bi-LSTM) and a Conditional Random Forest (CRF) layer to identify the CPE labels, such as vendor and product names, in the CVE summaries. The labels are then combined to reconstruct the CPE names. This method is like the approach we use in our paper. However, Wareus E. et al. \cite{Wareus} only considers the top 80\% of the most common CPE products and vendors names to "avoid CPEs with very few mentions". Those CPEs with very few mentions are eliminated. The disadvantage of this method is that it discourages the prediction of those rare CPE entities names in new CVE summaries, thus compromises the generality of the model.

Despite their efforts to extract CPEs from CVE summaries, achieving a high performance result is challenging. Many of these works fail to produce an accurate and precise CPE identification result. The model used in these researches are not the SoTA NLP model and their performance metrics are not good. Their F1 scores are below 90\%. Also, prior researches are not able to find and annotate enough corpora when training the model. Those proposed data annotation rules, such as the Database matching, Relevant Terms Gazetteer, and Heuristic Rules, are not able to provide an accurate annotation result. Hence, we draw insights from the challenges and advantages of the related works and design a system that addresses those disadvantages. Our proposed approach uses the state-of-the-art (SoTA) natural language processing (NLP) technique to automatically find CPE identities from new CVE summaries. We generated and annotated our training data using the SoTA Transformers models with two publicly available cybersecurity NER datasets. We also trained and evaluated the three SoTA Deep Learning NER models – BERT, XLNet, and GPT-2, to perform CPEs entity identification, and found that they outperformed previous methods \cite{Wareus} for automatic CPE labelling. The best model is selected to construct the CPE-Identifier system.

\section{Research Methodology}
\label{sec:Research Methodology}
\subsection{System Design}
\begin{figure}[]
	\centering
	\includegraphics[width=6cm]{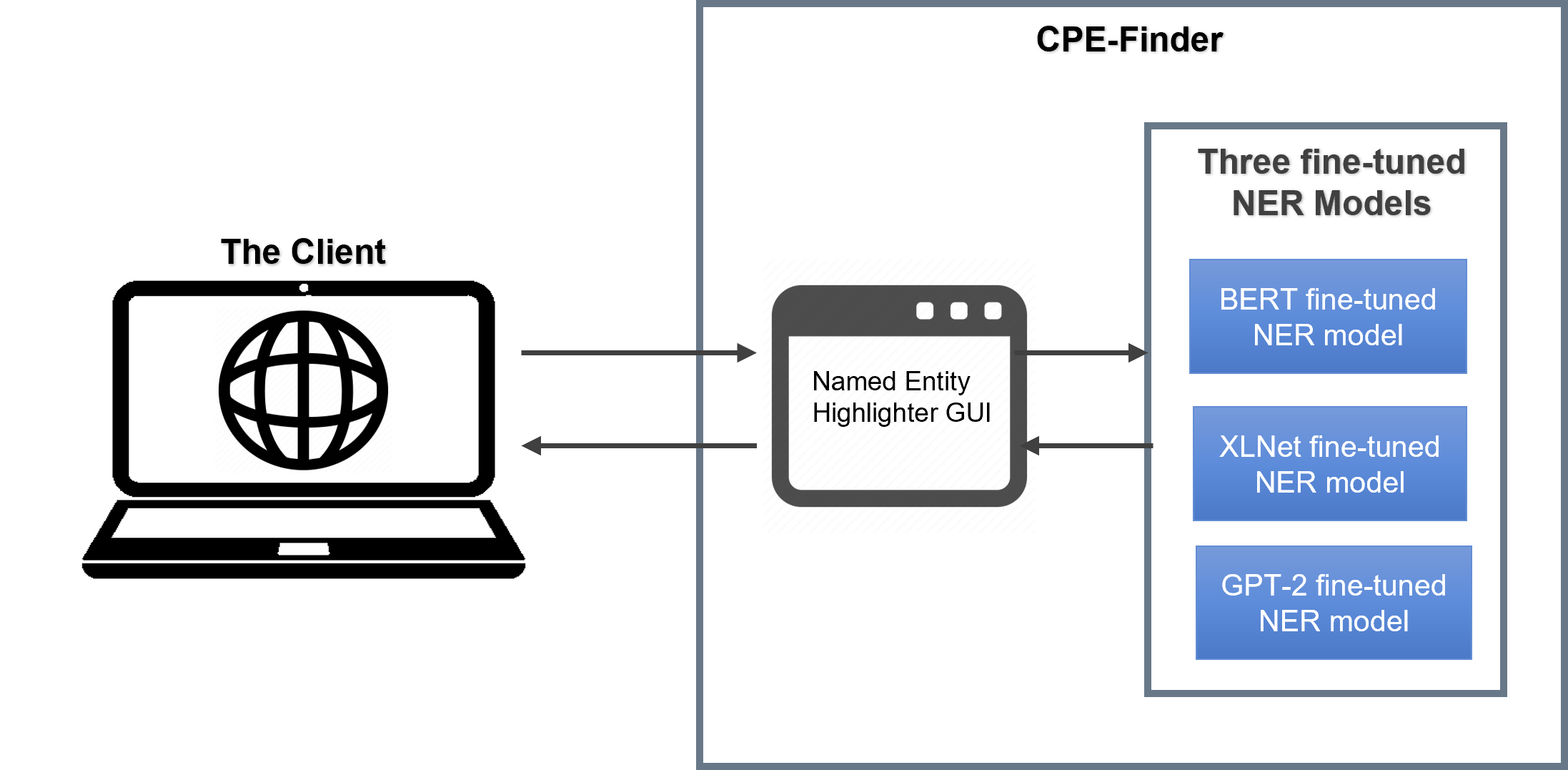}
	\caption{CPE-Identifier Design}
	\label{system_archi}
\end{figure}

The system architecture of the CPE-Identifier is a client and server model. The client  interacts with the CPE-Identifier using the Named Entity Highlighter GUI. After a string of CVE text is provided to the GUI, the text is encoded, supplied to the model, and then decoded to the prediction result. The Named Entity Highlighter GUI highlights the named entities and displays them to the client through the GUI. The Figure \ref{system_archi} illustrates the proposed CPE-Identifier design and its component.

\subsection{Proposed Approach}
\begin{figure}[]
	\includegraphics[width=7cm]{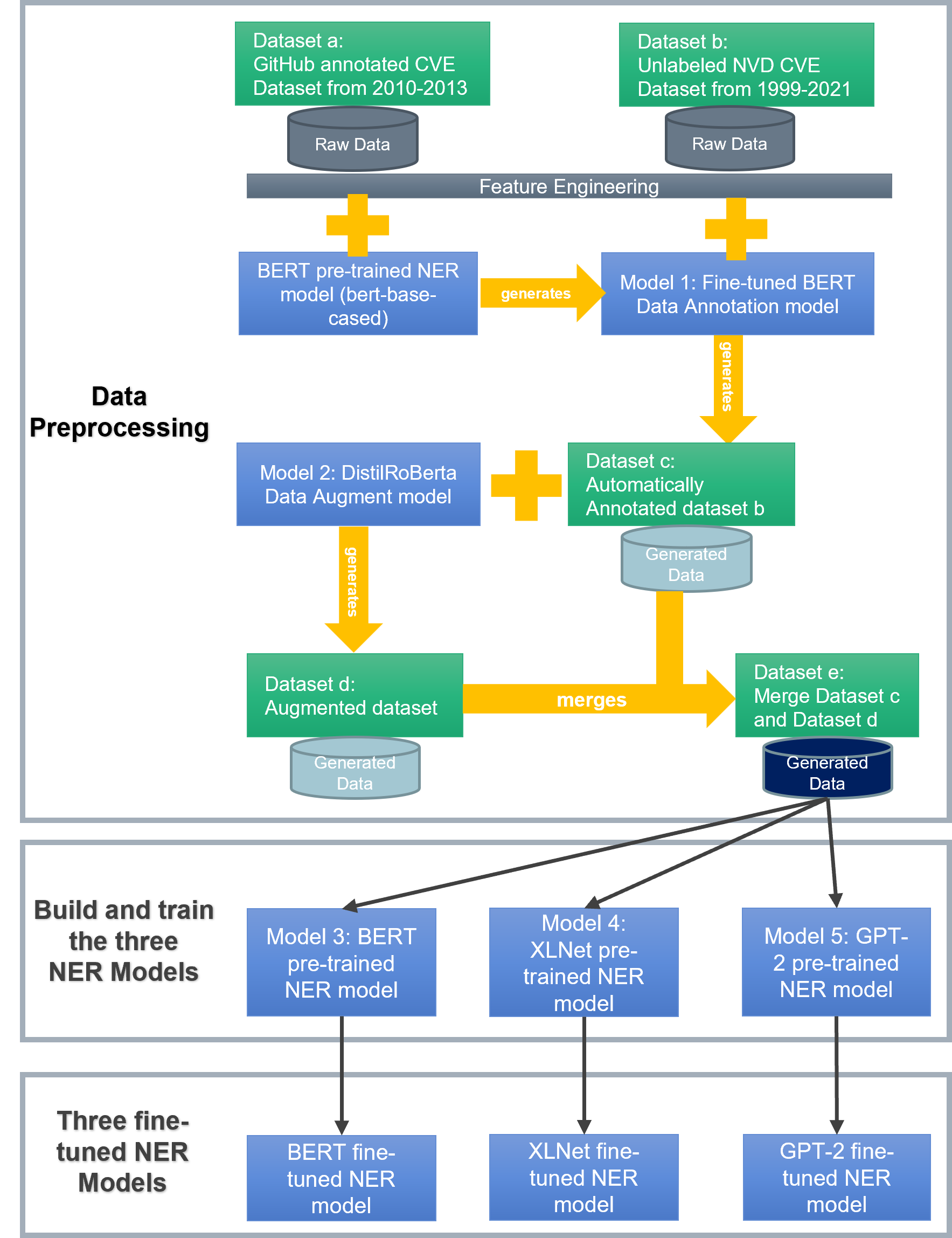}
	\centering
	\caption{Models Training Process Design}
	\label{model_archi}
\end{figure}

The Figure \ref{model_archi} presents the design of the models training process. The research consists of three stages:
\begin{enumerate}
	\item Data Pre-processing stage: This stage includes data collection, cleaning, annotation, augmentation, and merging. The CVE dataset is downloaded and undergone feature engineering processes. The result is used to generate new dataset and annotate raw dataset in this stage.
	\item Build and train models stage: This stage consists of constructing one Data Annotator model, one Data Augmentator model, as well as three SoTA NER models – BERT, XLNet, and GPT-2, which are trained using the pre-processed data.
	\item Analysis and build Graphical User Interface (GUI) stage: The predictive performance of each model are compared. A GUI is constructed to visualize the prediction result of the best model.
\end{enumerate}

\section{Dataset and Labels}
\label{sec:Dataset and Labels}
To fine-tune a model capable of understanding cybersecurity domain-specific jargon, we need a large set of labeled data specifically for the Named Entity Recognition (NER) task that are tailored to the cybersecurity domain. Moreover, high-quality annotated entities closely related to the CVE-CPE field are also important for identifying new entities in this domain. In a recent paper \cite{Aghaei}, Aghaei et al. also suggested that the absence of publicly available domain-specific NER labeled dataset has caused their research on training a NER model in cybersecurity a "challenging task". The author employed a relatively small sized dataset called MalwareTextDB \cite{MalwareTextDB}. However, after investigating the MalwareTextDB dataset, it consists of four NER tags, which are not related to our research target. Furthermore, the size of the dataset is relatively small for training a deep learning model, which may cause the model to overfit. Therefore, it is crucial to collect high-quality publicly available training datasets exclusively compiled for the cybersecurity CVE-CPE domain.

However, large and annotated CVE-CPE datasets are not publicly available. A few possible reasons for this challenge include:

\begin{itemize}
	\item \textbf{Challenge 1}: There are few publicly available annotated CVE datasets for NER tasks.
	
	Compared to other natural language processing tasks such as text classification or relation extraction, Named Entity Recognition is not common on CVE datasets. There is only a few past research works focused on the NER task of the CVE dataset, making it even more challenging to find a large amount of annotated NER corpus. To the best of our knowledge, there is only one such publicly available dataset from the Stucco project \cite{GitHub_data}.
	
	\item \textbf{Challenge 2}: Those publicly available annotated CVE datasets are not the latest. 
	
	The CVE summary and its corresponding CPE metadata are updated every day. It is difficult for the cybersecurity analysts to keep an eye on every new CVE released and maintain an updated database. Moreover, the CVE NER dataset is not used by any specific application or computing environment but only by developers. Therefore, there is no complete set of the latest annotated CVE NER dataset.
	
	\item \textbf{Challenge 3}: Manual annotation of NER training data is costly.
	
	The manual annotation of a high-quality CVE dataset for Named Entity Recognition in the cybersecurity domain is time-consuming, requiring experts who are professionals in cybersecurity.
	
\end{itemize}

Training a model using a small amount of data or a highly skewed dataset will cause the model to over-fit. The model “sees” a small number of tokens repeatedly and learns to recognize those training data correctly but is not capable of generalizing to the novel inputs in the test dataset. The training error of the model becomes extremely small at the expense of high test-error.

The figure \ref{cs} shows our proposed solution to each challenge of the data availability:
\begin{figure}[]
	\centering
	\includegraphics[width=6cm]{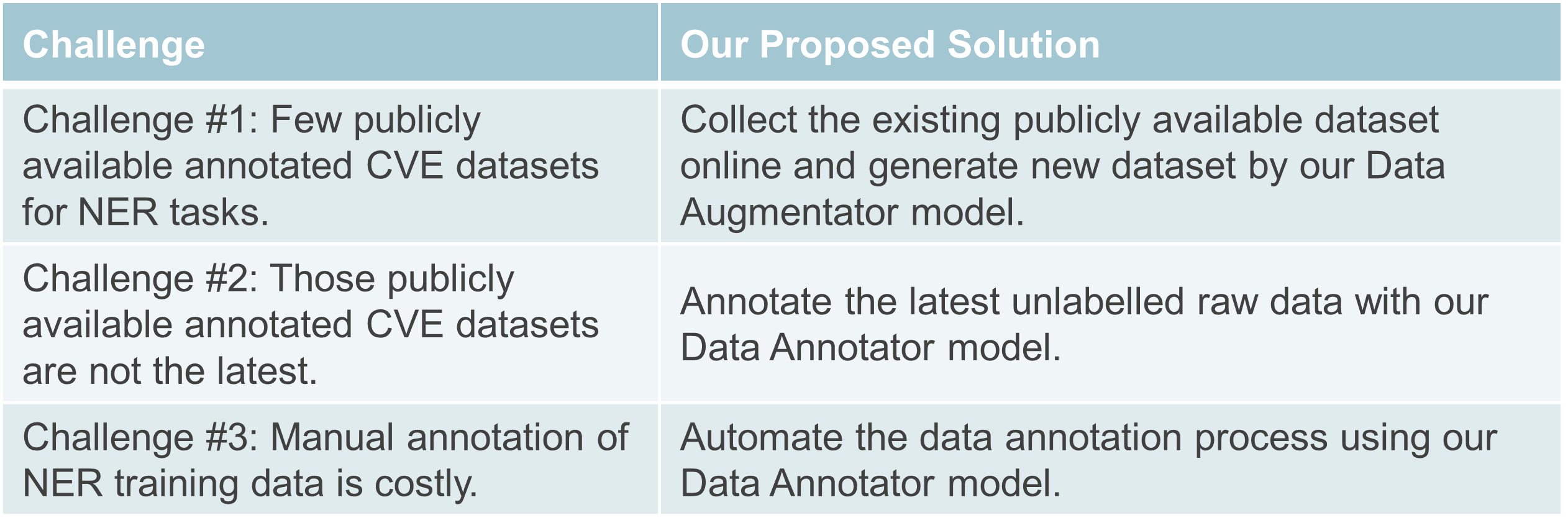}
	\caption{Challenges and Proposed Solutions}
	\label{cs}
\end{figure}

\subsection{CPE data}
CPE is a standardized identification system for Information and Communications Technology (ICT) products. It provide a more consistent and accurate way to identify products across different security databases, and make it easier for security researchers to collect and share information about vulnerabilities and products. Each CPE is linked to a particular CVE ID, which includes the following features \cite{Andrew}: \textit{cpe\_version, part, vendor, product, version, update, edition, and language}. The CPE texts are available at the NVD official website\cite{NVD_Feed_data}.

\subsection{Padding/Trimming the sentences}
Since each of the sentences in the dataset is of a different length, we have padded and trimmed the input sentences to keep all the sentences to be the same length. 

Out of 361088 sentences, the length ranges from 3 to 449 words. There are 93.96\% of the sentences are shorter than 128 words, therefore we have set the maximum length of the sentences to 128 tokens. Those sentences shorter than 128 are padded with [PAD] tokens. Those sentences longer than 128 are truncated.
\begin{figure}[]
	\centering
	\includegraphics[width=5cm]{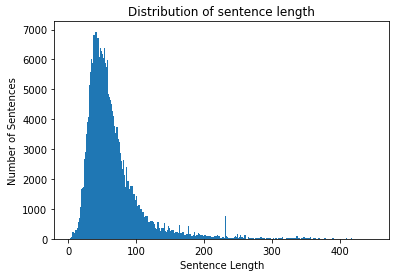}
	\caption{Sentence Length Distribution}
	\label{sentence_distribution}
\end{figure}	

\subsection{Tagging Schemes}
The training corpora are required to be tagged before performing training. The tagging scheme, BIO (Beginning, Inside, Outside), is used in this paper.

%
%
%
%
%
%

\section{Models}
\label{sec:Models}
We trained and used five pre-trained models and one Masked-Language Model (MLM) to construct our CPE-Identifier system. These six models are:

\begin{enumerate}
	\item \textbf{BERT pre-trained NER model} 
	\begin{itemize}
		\item This is a pre-trained bert-base-cased model downloaded from huggingface.co. It is used to fine-tune our own Data Annotator model from scratch.
	\end{itemize}
	\item \textbf{Model 1: Finetuned BERT Data Annotator model}
	\begin{itemize}
		\item This is a fine-tuned bert-base-cased model from training annotated cybersecurity NER texts from GitHub (Dataset a). This model is used to annotate the named entities in NVD CVE Dataset from 1999-2021 (Dataset B) and produces an Automatically Annotated dataset (Dataset C).
	\end{itemize}
	\item \textbf{Model 2: DistilRoBerta Data Augmentator model}
	\begin{itemize}
		\item This is a distilroberta-base Masked Language model. This model generates the Augmented NER dataset (Dataset D) by randomly replacing word tokens in sentences from the Automatically Annotated dataset (Dataset C).
	\end{itemize}
	\item \textbf{Model 3: BERT NER model}
	\begin{itemize}
		\item This is a pre-trained bert-base-cased model.
		\item We use the Final Training Corpora (Dataset E) to train this model and analyze its performance against Model 4 and Model 5.
	\end{itemize}
	\item \textbf{Model 4: XLNet NER model}
	\begin{itemize}
		\item This is a pre-trained xlnet-base-cased model.
		\item We use Final Training Corpora (Dataset E) to train this model and analyze its performance against Model 3 and Model 5.
	\end{itemize}
	\item \textbf{Model 5: GPT-2 NER model}
	\begin{itemize}
		\item This is a pre-trained gpt2 model.
		\item We use Final Training Corpora (Dataset E) to train this model and analyze its performance against Model 3 and Model 4.
	\end{itemize}
\end{enumerate}

\subsection{\textbf{BERT model}}

BERT is an AutoEncoder (AE) language model and uses only the Encoder portion of the Transformer architecture. We construced our Data Augmentator model by utilizing the BERT MLM property to generate new sentences. We masked seven tokens in each sentence and asked the model to predict the masked word.

The advantage of the AE language model is that it can see the context in both the forward and backward directions. However, it uses the [MASK] token in the pre-training, which is absent from the real data during fine-tuning and results in the pretrain-finetune discrepancy. It also assumes the masked token is independent of unmasked tokens, which is not always the case. For example, “Microsoft Word” where “Word” is dependent on “Microsoft”. If “Word” is masked and predicted, Bert may output “Microsoft Office”. Therefore, we also tested the XLNet model, which addresses this issue.

\subsection{\textbf{XLNet model}}
The XLNet model is a generalized Auto-Regressive (AR) Permutation Language model \cite{Liang}. It is a bi-directional Language model. The AR model can predict the next word using the context word from either forward or backward, but not both directions.

\begin{figure}[]
	\centering
	\includegraphics[width=5cm]{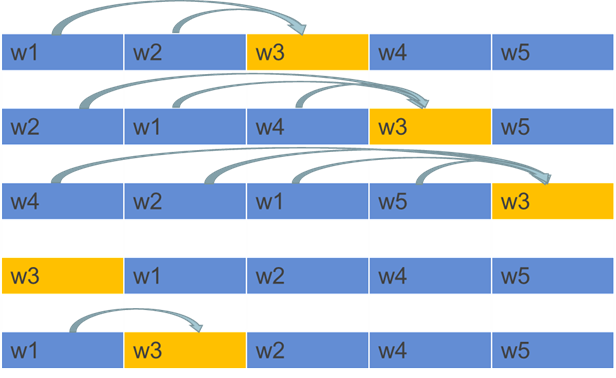}
	\caption{XLNet Permutation Language Modelling (PLM)}
	\label{xlnet_model}
\end{figure}

XLNet solved this issue by introducing Permutation Language Modelling (PLM), PLM allows for predicting the upcoming words by permutating a sequence and gathering information from all positions on both sides of the token. In the figure, the target token w3 is located at the i-th position, and the i-1 tokens are the context words for predicting w3. There are five patterns in 120 permutations. The words are permutated at each iteration; therefore, every word will be used as the context word. This way, it can avoid the disadvantages of the MASK method in the AE language model. 

\subsection{\textbf{GPT-2 model}}
GPT-2 is Auto-Regressive (AR) model as well \cite{Liang}. It is good at generative NLP tasks because of its forward direction property. However, it can only use forward or backward contexts, which means it can't use forward and backward context simultaneously. Therefore, GPT-2 is a Uni-directional Language Model, unlike the other two NLP models. Different from the Bert model, GPT-2 uses the Decoder portion of the Transformers architecture. 

\section{Implementation}
\label{sec:Implementation}
The implementation of the research follows three stages: Data pre-processing, models training, and design and development of the  Graphical User Interface (GUI).

\subsection{Data Pre-processing Stage}
This stage aims to create a set of clean, annotated, and structured NER corpora, which will be used to train the three SoTA models.

We identified five datasets to be used for our work. They are:
\begin{enumerate}
	\item Dataset A: GitHub annotated CVE Dataset from Jan 2010 to March 2013 publicly available online \cite{GitHub_data}.
	\item Dataset B: Raw CVE Dataset from 1999-2021 publicly available online \cite{NVD_Feed_data}.
	\item Dataset C: Self-annotated NER dataset using our proposed Data Annotator model.
	\item Dataset D: Self-augmented NER dataset using our proposed distilroberta-base Masked Language model.
	\item Dataset E: Final Training Corpora generated by combining both Dataset C and Dataset D, which is used to train the three SoTA models.
\end{enumerate}

\begin{figure}[]
	\centering
	\includegraphics[width=6.5cm]{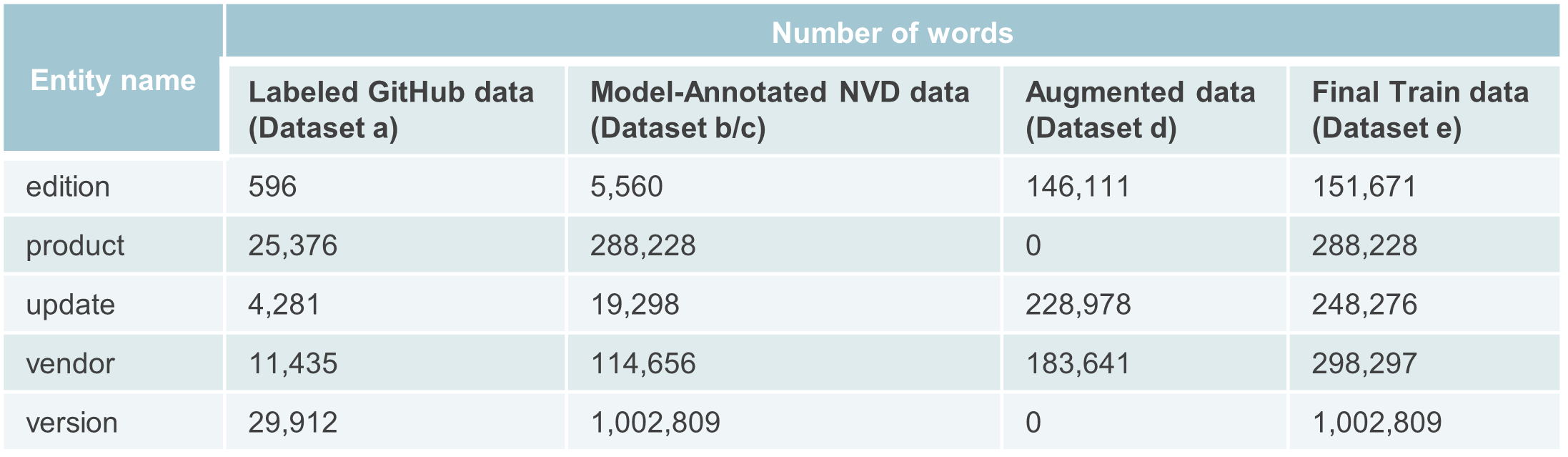}
	\caption{The details of all Datasets used and generated}
	\label{dataset_all}
\end{figure}

\subsubsection{Data Collection}
Dataset A is downloaded from GitHub \cite{GitHub_data}. It consists of CVE summaries from three sources: Microsoft Security Bulletin, Metasploit Framework database, and the NVD CVE database. The entity types of each dataset is shown in figure \ref{datasetA}:

\begin{figure}[]
	\centering
	\includegraphics[width=6.5cm]{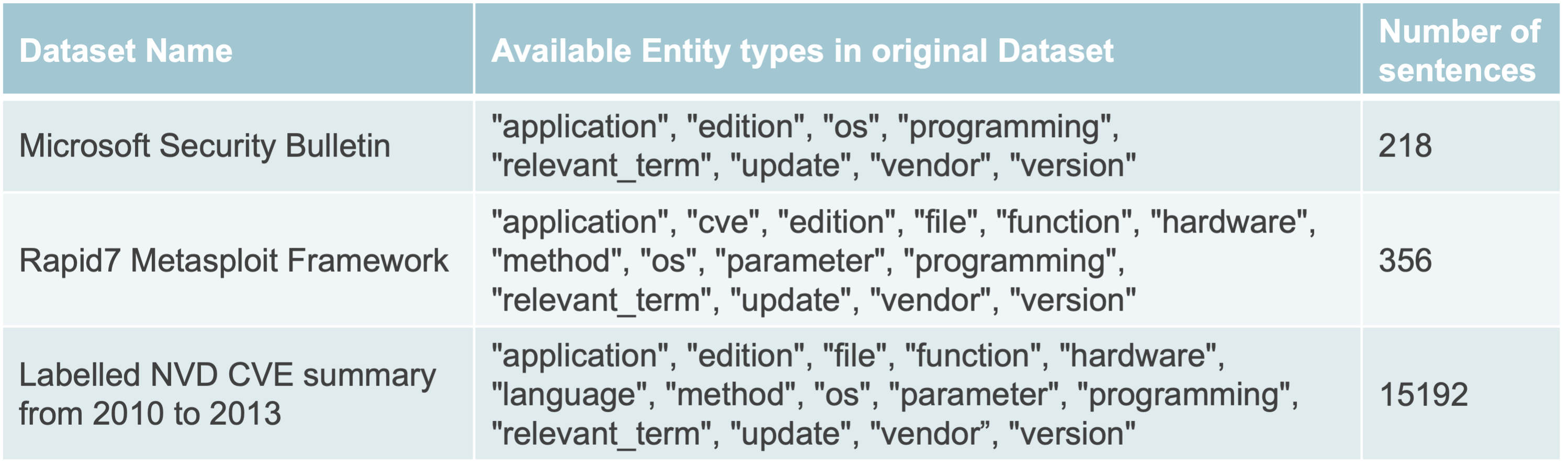}
	\caption{Dataset A}
	\label{datasetA}
\end{figure}

While the same CVE-ID describes the same kind of vulnerabilities, the descriptions in each source are different. For example, in Figure \ref{cve_texts} the descriptions of CVE-ID: CVE-2013-1293 are different from various sources:

\begin{figure}[]
	\centering
	\includegraphics[width=6.5cm]{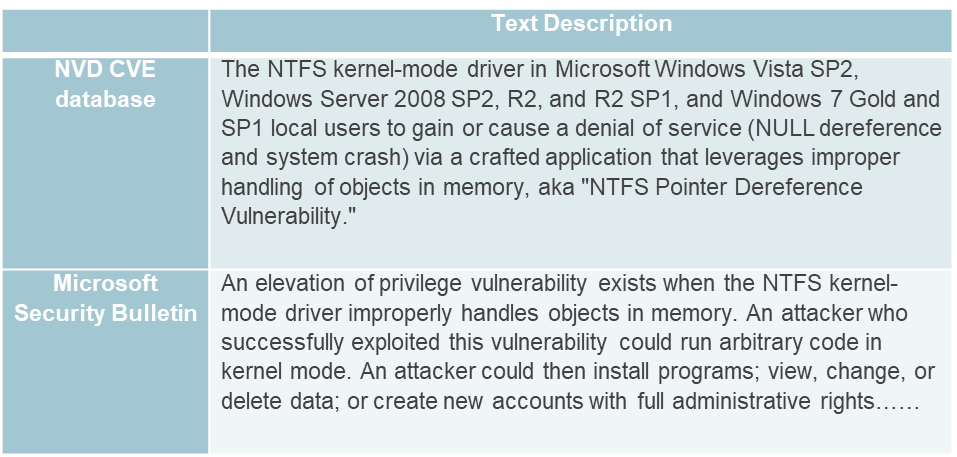}
	\caption{CVE text Description from different sources}
	\label{cve_texts}
\end{figure}

Dataset B contains the raw CVE summaries and CPEs from 1999 to 2021 that is downloaded from \cite{NVD_Feed_data}.

\subsubsection{Data Cleaning}
Our proposed CPE-Identifier system only considers the top five popular named entities in the CPEs -- “edition", "product", "update", "vendor", and "version“, because the remaining named entities are not always available in the CPE identifier strings dataset. These five types are enough for the cybersecurity analyst to quickly narrow down to the vulnerability details. Therefore, We have replaced the unrelated entities in dataset A with the Out-of-Word label ("O"). After investigating the dataset, we have noted that the "application", "hardware" and "os" labels correspond to the "product" label in the CPE identifier strings, so we have renamed the three entities to "product".

We then constructed two 2-dimensional lists to store the data in NER format. The two sample 2-dimensional lists to store the data in NER format is illustrated \ref{NER_data_format}:

\begin{figure}[]
	\centering
	\includegraphics[width=7cm]{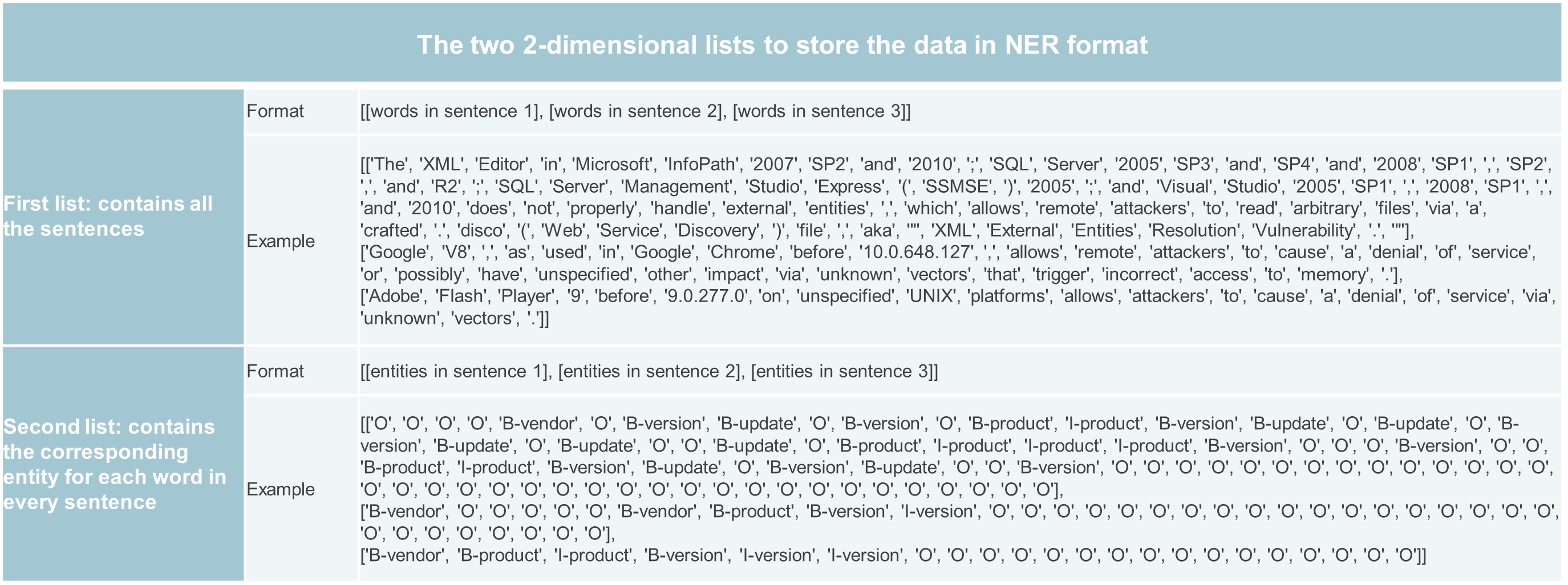}
	\caption{Sample data format after Data Cleaning step}
	\label{NER_data_format}
\end{figure}

For dataset B, we combined the dataset from 1999 to 2021 into one text file.

\subsubsection{Data Annotation}
This step allows an NLP model to automate the named entity labeling process. Manually annotating the 169092 unlabelled CVE summaries is time-consuming and costly and requires expert knowledge in both cybersecurity and natural language processing. Therefore, we built a deep learning model to annotate the unlabelled CVE summaries automatically. It dramatically improves efficiency and accuracy of the data labeling process. The model accuracy is 98.73\%, and its F1 score is 95.64\%. The time taken to annotate 169092 sentences is only 2 hours. The high F1 score as well as the accuracy score proves that our fine-tuned BERT Data Annotator model is capable of tagging the sentences with NER tags accurately. Furthermore, we have also randomly picked sentences from the output dataset (dataset C) and manually examined its accuracy. The steps to generate and annotate the dataset C is described.

\begin{enumerate}
	\item Constructed the Data Annotator model by finetuning the bert-base-cased NER model from scratch using dataset A. The resulting model (model 1) learns the domain knowledge from cybersecurity texts and will be used as a tool in the next step.
	\item We used model 1 as an annotating tool to identify the named entities in the unlabelled NVD CVE dataset (dataset B) from the Data Cleaning stage. 
	\item The resulting labeled data (datast C) contains 169092 sentences. The distribution of the entities is shown in the figure \ref{dataset_all}.
\end{enumerate}

\subsubsection{Data Augmentation}
The next step in the Data Pre-processing stage is Data Augmentation. The augmented dataset was generated using the NVD CVE summary with BERT Masked Language Model. Data augmentation helps to improve the data skewness problem by increasing the data size of the specific named entity. For example, in our automatically annotated NVD CVE summary dataset, there are only 19298 tokens that are tagged as the “update” entity, while 288228 words are classified as the “product” entity. This highly imbalanced data causes the resulting model to be less generalized on the “edition” tag.

We build a BERT Masked Language Model and augment new sentences by randomly replacing seven words with their synonyms based on the dictionary. There are very few “edition”, “vendor” and “update” named entities in our automatically annotated NVD CVE summary dataset – 5560 words for “edition, 114656 for “vendor” and 19298 for “update”; we plan to generate more sentences containing these three entities.

We extracted sentences containing “edition”, “vendor” and “update” only. There are a total of 13288 sentences. We created 192380 new sentences using our augmentation model, the Data Augmentor. The distribution of the augmented data is shown in the Figure \ref{dataset_all}.

\subsubsection{Data Merging}
The last step in the Data Pre-processing stage is to merge the Automatically Annotated NER dataset (dataset C) and the Augmented NER dataset (dataset D). The distribution of the entities is shown in the Figure \ref{dataset_all}. The Final Training Corpora contains 361472 sentences and will be used to train the three SoTA NER models.

\subsection{Models Training Stage}
The models are trained in the Ubuntu Kernel, using GeForce RTX 3070 Graphics Card. CUDA version 11.3. 

\subsubsection{\textbf{Model 1: Finetuned BERT Data Annotator model}}
The fine-tuned BERT Data Annotator model is fine-tuned from the bert-base-cased pre-trained model for 5 epochs, using the annotated CVE summaries (dataset a) from GitHub. The purpose of the Data Annotator model is to serve as a tool to annotate the unlabelled NVD CVE summaries (Dataset B) with the named entities.

The algorithm of the approach is shown in Fig. \ref{algo_1}:

\begin{figure}[]
	\centering
	\includegraphics[width=6cm]{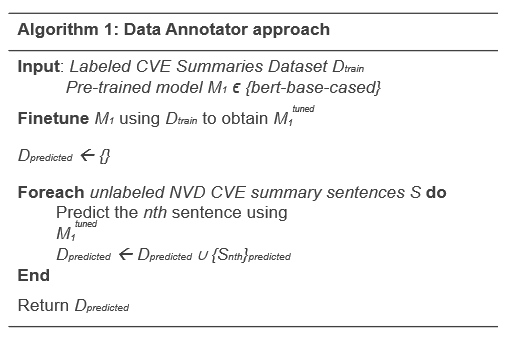}
	\caption{Data Annotator Approach}
	\label{algo_1}
\end{figure}

\begin{figure}[]
	\centering
	\includegraphics[width=6cm]{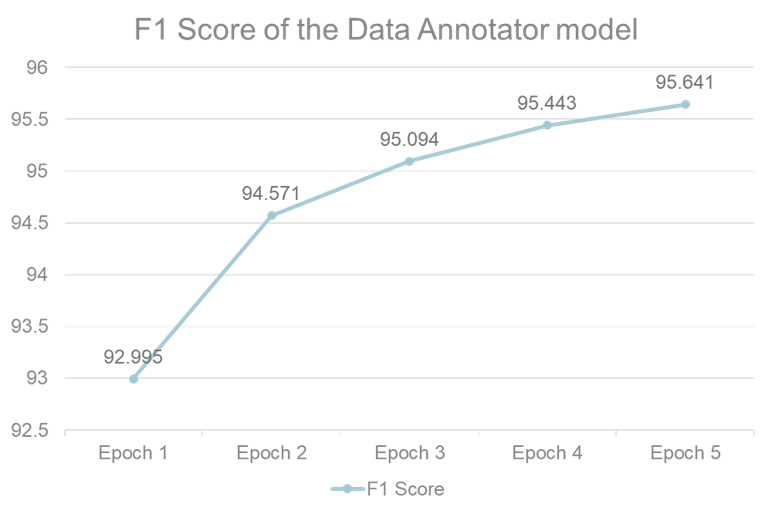}
	\caption{F1 score of Data Annotator model}
	\label{annotator_f1_graph}
\end{figure}

The F1 score of the model is 95.64\%, and the accuracy score is 98.72\%. The F1 score is plotted in Fig. \ref{annotator_f1_graph}. This approach improves efficiency and saves the cost of the manual labeling process.

\subsubsection{\textbf{Model 2: DistilRoBerta Data Augmentor model}}
The DistilRoBerta Data Augment model is built on top of the DistilRoBerta \cite{Distilroberta} pre-trained model. The purpose of the Data Augmentor model is to serve as a tool to generate new labeled NER corpora. The imbalanced data causes skewness issue in the model, which degrades the model to describe typical cases as it must deal with rare cases on extreme values. The model will predict better on entities with more training words than those with fewer training words \cite{C.}. This approach solves the data skewness issue by generating new sentences with specific entities. The algorithm of the approach is shown in \ref{algo_2}:
\begin{figure}[]
	\centering
	\includegraphics[width=5cm]{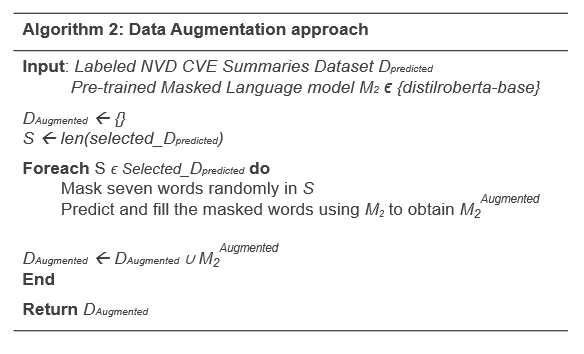}
	\caption{Data Augmentation Approach}
	\label{algo_2}
\end{figure}

\subsubsection{\textbf{Model 3, 4, 5: BERT, XLNet and GPT-2 NER models}}
The three SoTA NER models are trained for 20 epochs each. The figure \ref{training_parameters} shows the hyperparameters used when training the models:
\begin{figure}[]
	\centering
	\includegraphics[width=5.5cm]{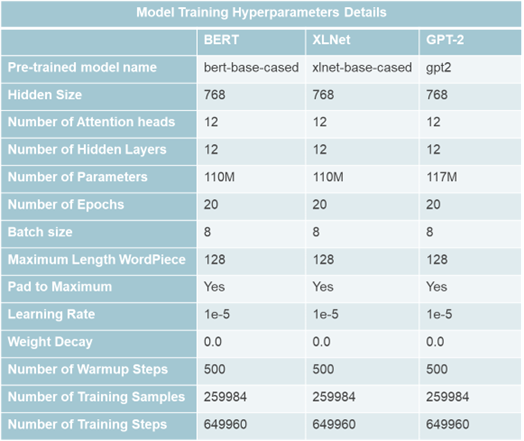}
	\caption{Model Training Hyperparameters Details}
	\label{training_parameters}
\end{figure}

We have tested the BERT, XLNet and GPT-2 models and compared their performances and determined the most suitable architecture for the NER task on CVE summaries.

\subsection{Analysis and build Graphical User Interface (GUI)}
We have also designed a GUI to visualize the prediction result. The interface of the CPE-Identifier is hosted on a web using the Streamlit Python package. The Interface allows the analysts to input the new CVE texts into the “Enter Text Here” text box and select a specific model, or all the three fine-tuned models to retrieve the annotated result. This feature allows the analysts to choose the best model that fits their CVE text.

\begin{figure}[]
	\centering
	\includegraphics[width=1.5cm]{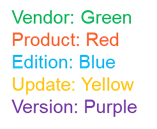}
	\caption{CPE Entities Colors}
	\label{colors}
\end{figure}

The prediction result is highlighted in different colors, where each color indicates a named entity type. The colors and their corresponding entities are shown in Figure \ref{colors}

The figure \ref{GUI} shows the input text box and the corresponding prediction result, where entities are highlighted in the corresponding colors:

\begin{figure}[]
	\centering
	\includegraphics[width=7cm]{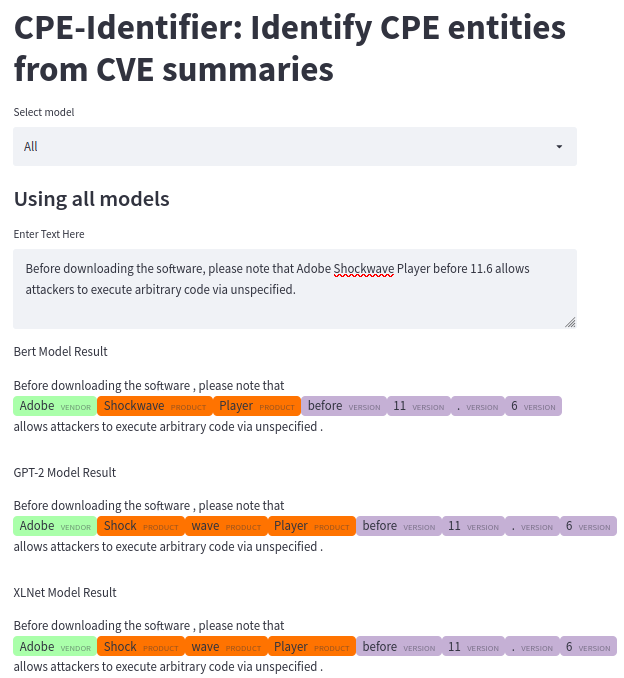}
	\caption{GUI of CPE-Identifier}
	\label{GUI}
\end{figure}

In the example, the word "Adobe" is highlighted in green, indicating it belongs to the "VENDOR" category, The word "Shockwave Player" is in orange, which means that it is part of "PRODUCT" category. One interesting finding is that, instead of one word corresponds to one entity, the same word can be identified as different entities at different positions of the sentence.
For example, according to dictionary.com\cite{dict}, the word "before" is defined as "previous to the time when". It does not belong to any entities when used alone. However, in the example sentence, the word "before" is identified as different entities when it is positioned differently.
The word "before" at the beginning of the sentence is not highlighted, which indicates that it is not an entity. However, the word "before" in the front of the word "11.6" is highlighted in purple, which indicates that it belongs to the "VERSION" entity. This finding shows that our fine-tuned BERT model is intelligent enough to identify the meanings of the words based on their syntactic and semantic positions in the sentence, and not purely based on the definitions of that particular words. This is a good indicator of the performance of our fine-tuned model.

\section{Result Analysis}
\label{sec:Result Analysis}
The models’ performances are evaluated using Accuracy, Precision, Recall, and the F1 score and training time. The results are shown in the table.

\begin{figure}[]
	\centering
	\includegraphics[width=5cm]{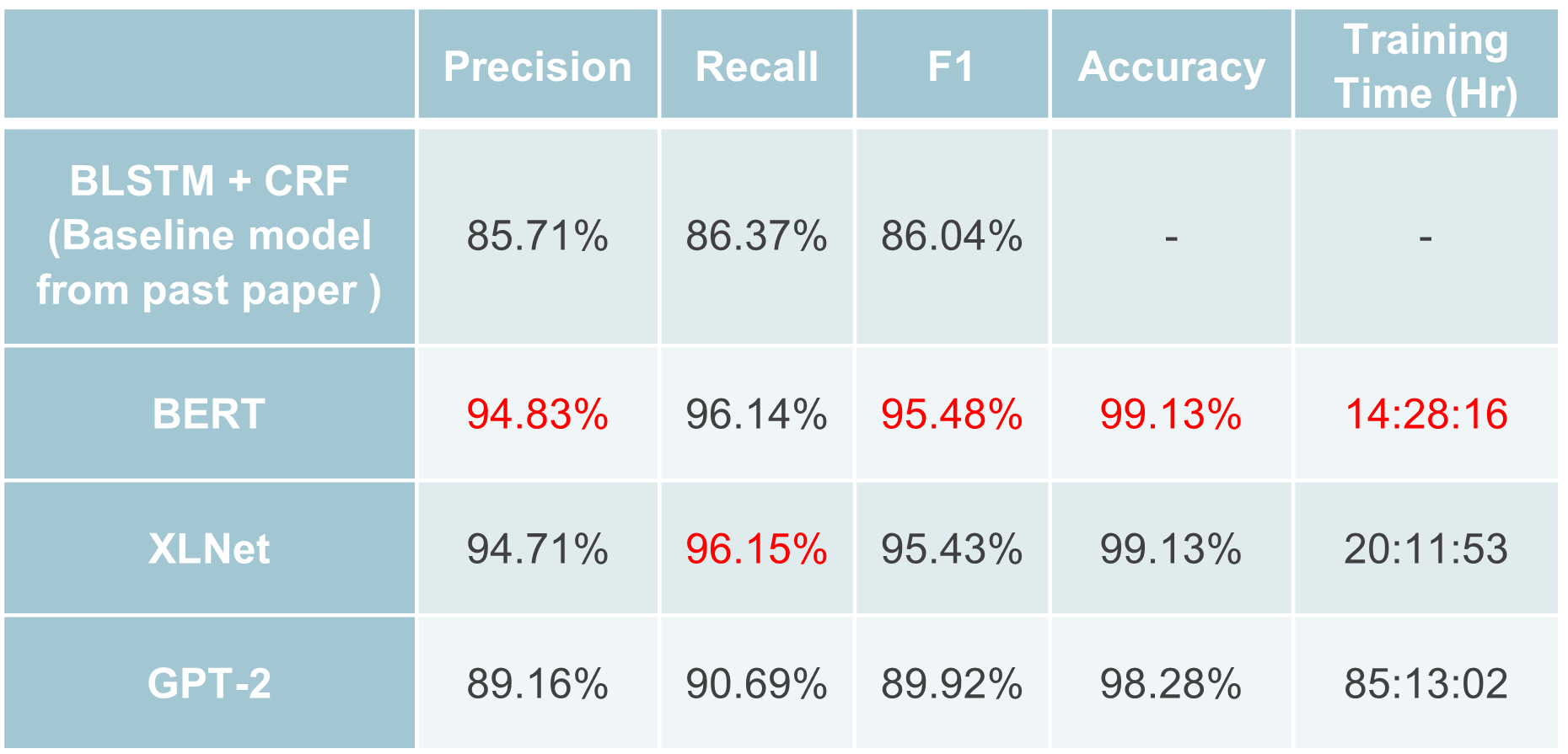}
	\caption{Model Results Comparison}
	\label{results}
\end{figure}
\subsection{Precision}
The formula to calculate the precision score is
\[Precision\_Score=\frac{True Positive}{True Positive+False Positive}\]
The BERT model achieved the highest Precision score, with a value of 94.83\%. XLNet gained 94.71\%, and GPT-2 obtained 89.16\%.
\begin{figure}[]
	\centering
	\includegraphics[width=5cm]{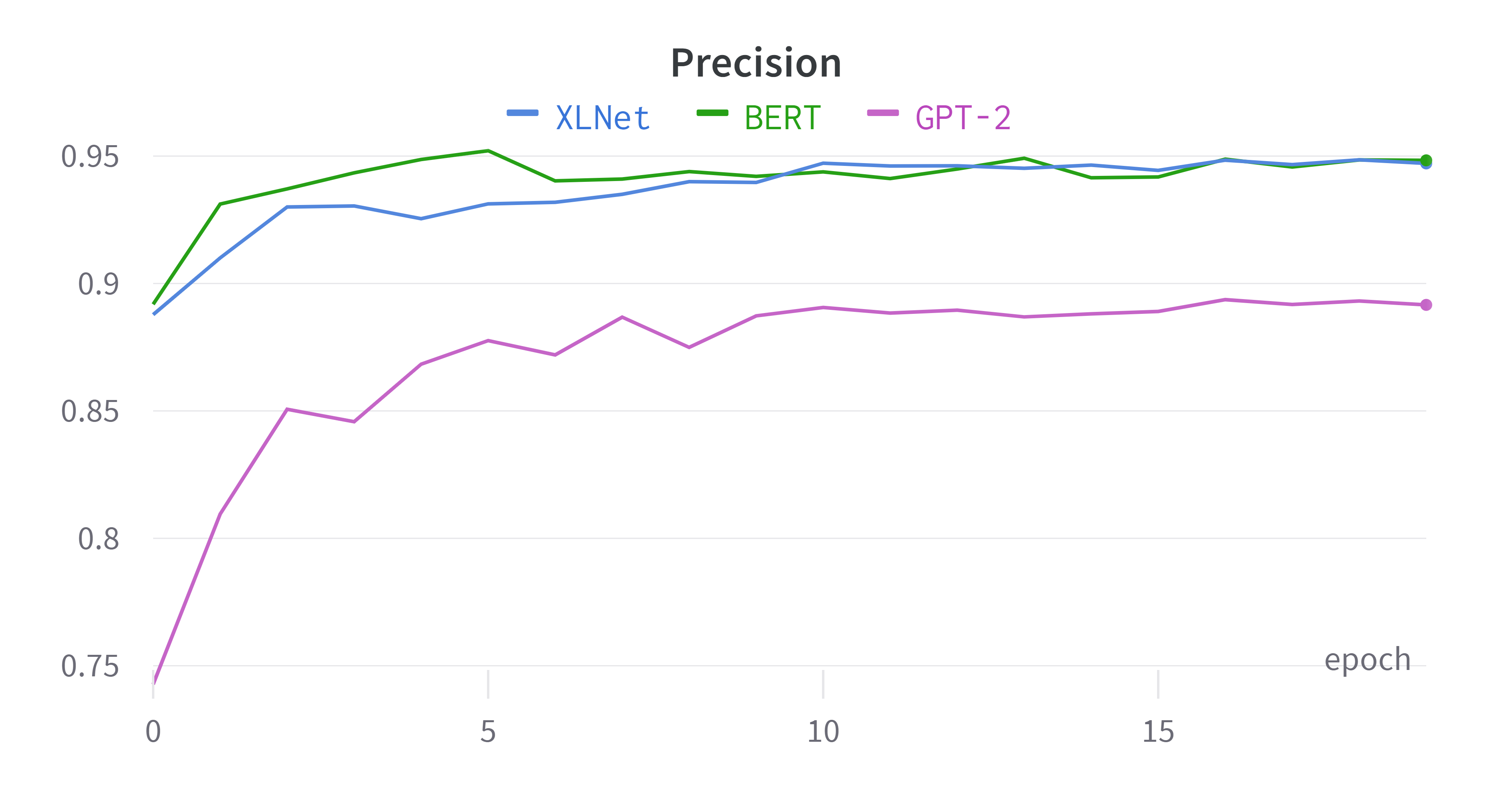}
	\caption{Model Precision Comparison}
	\label{precision}
\end{figure}

\subsection{Recall}
The formula to calculate the recall score is
\[Recall\_Score=\frac{True Positive}{True Positive+False Negative}\]
The XLNet model achieved the highest Recall score, with a value of 96.15\%. Bert gained 96.14\%, and GPT-2 obtained 90.69\%.
\begin{figure}[]
	\centering
	\includegraphics[width=5cm]{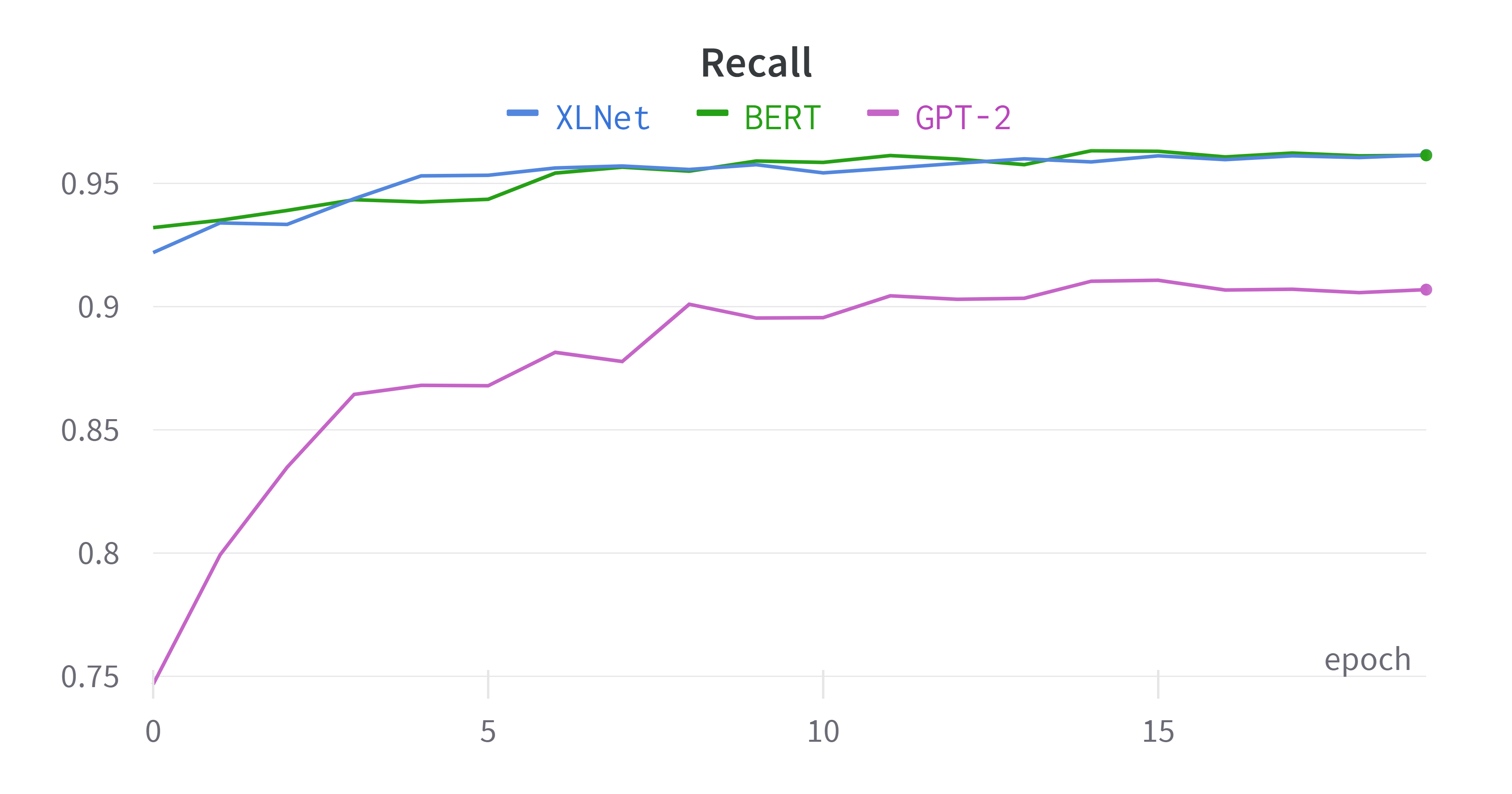}
	\caption{Model Recall Comparison}
	\label{Recall}
\end{figure}

\subsection{Accuracy}
A model with a high accuracy means it can locate most of the named entities correctly. The formula to calculate accuracy is
\[Accuracy\_Score=\frac{TP+TN}{TP+FP+TN+FN}\]
The BERT and XLNet models achieved the highest Accuracy scores, with a value of 99.13\%, and GPT-2 obtained 98.28\%.

\begin{figure}[]
	\centering
	\includegraphics[width=5cm]{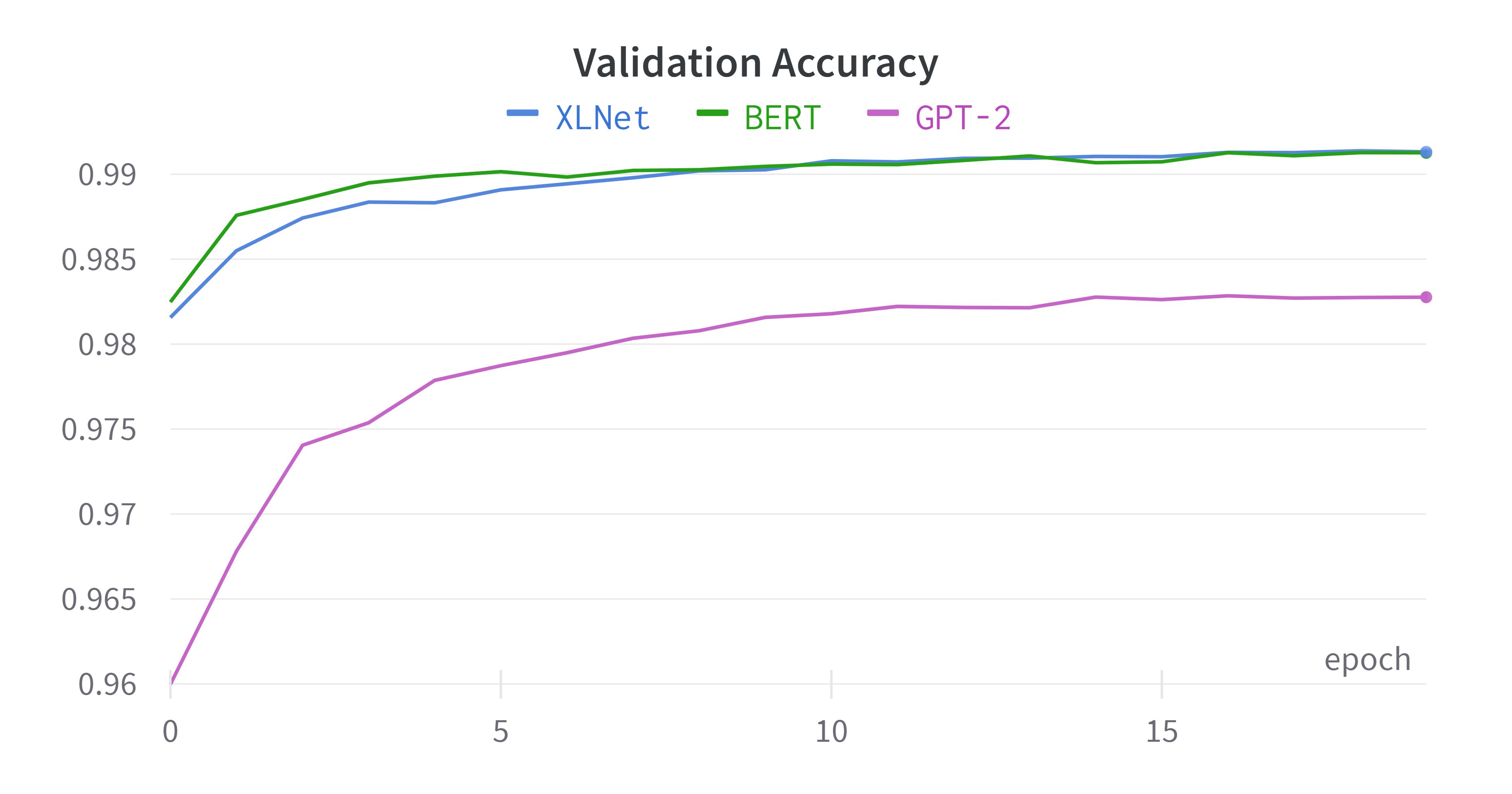}
	\caption{Model Accuracy Comparison}
	\label{Accuracy}
\end{figure}

However, although the model achieves an accuracy above 90\%, 10\% of the cases are the named identities, but the model predicts that they are not. This is obviously too high a cost for the cybersecurity expert to neglect that case. One missed named entity may cause the CPE-Identifier becomes unreliable and results in a single point of failure.  Therefore, we also take the F1 score into account.

\subsection{F1 score}
The F1 score is the most critical evaluation metric in machine learning. It considers precision and recall and gives a measure of the incorrectly identified cases. F1 score is a good assessment of the model as it considers the distribution of the data. If the data is highly imbalanced, the F1 score better assesses the model’s predictive performance.

If our model incorrectly predicts positive named entities as negative or vice versa, the cybersecurity analysts will get confused about the prediction result. It is crucial to ensure there is as few False Positive and False Negative as possible, which means both precision and recall are maximal. However, in practice, it is impossible to maximize the precision and recall together. Improving the precision score will reduce the recall score and vice versa. Therefore, we used the F1 score as it considers both precision and recall.

The F1 score is the weighted average of the model’s precision and recall. The range of the F1 score is between 0 and 1. 0 indicates the predictive performance of the model is worst, while 1 represents the predictive performance of the model is the best. The formula of the F1 score is 
\[F1=2*\frac{Precision*Recall}{Precision+Recall}\]	
The F1 score of the three models is plotted in a graph and compared. The BERT model achieved the highest F1 score, with a value of 95.48\%. XLNet gained 95.43\%, and GPT-2 obtained 89.92\%.
\begin{figure}[]
	\centering
	\includegraphics[width=5cm]{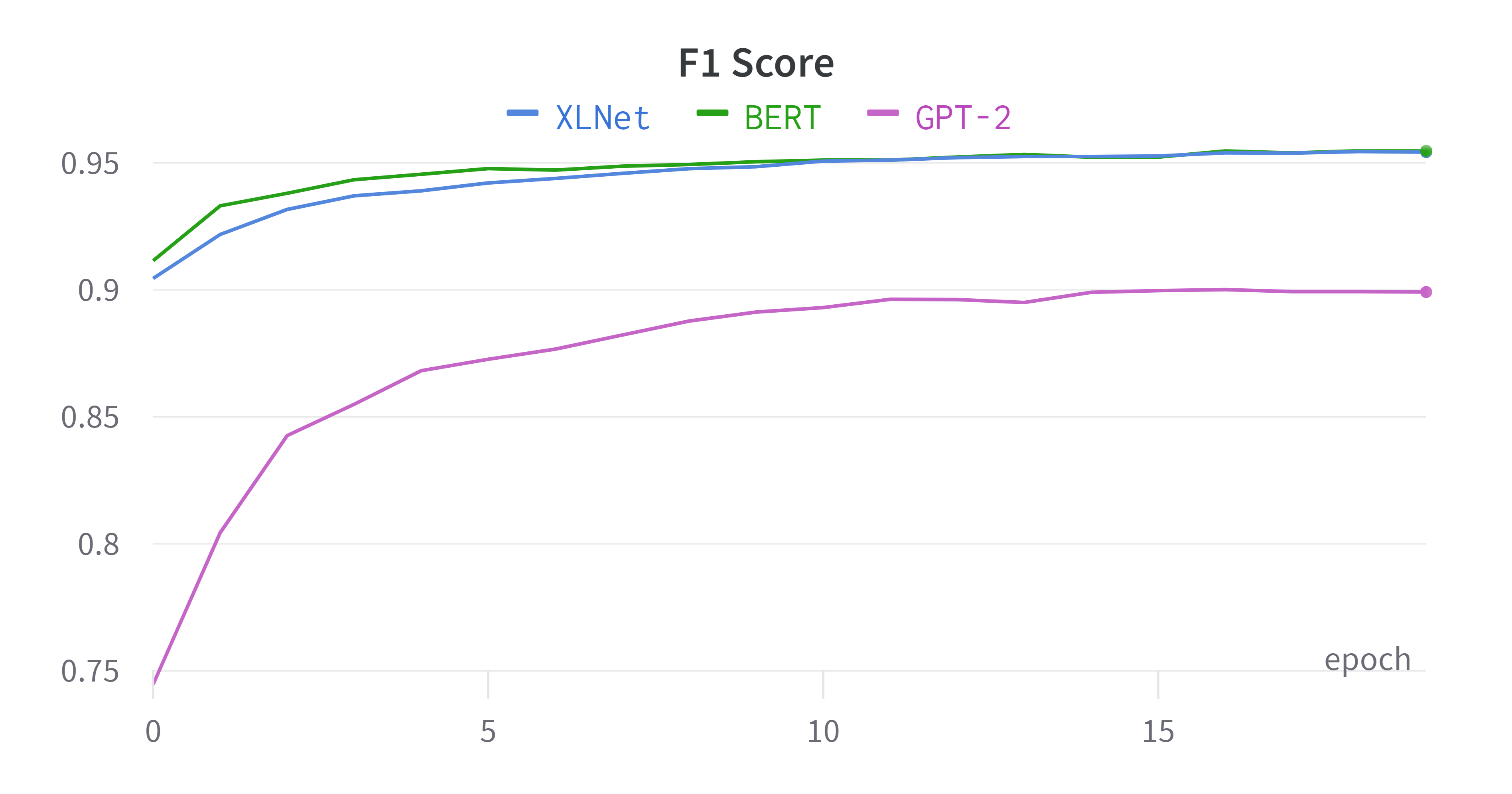}
	\caption{Model F1 Score Comparison}
	\label{f1}
\end{figure}
Hence, from the performance metrics above, the Bert model is the best model which obtained the highest Precision, Accuracy, and F1 score. Its Recall score is 0.01\% lower than the XLNet model.

\subsection{Error Analysis}
\begin{figure}[]
	\centering
	\includegraphics[width=7cm]{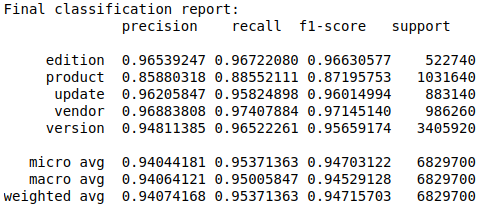}
	\caption{Classification Report}
	\label{classification_report}
\end{figure}

For the model with the best performance -- the Bert model, we have done an error analysis and checked the number of CVEs in the test set have been correctly labeled for CPE without any error. The classification report of the prediction result is as shown in the Figure \ref{classification_report}

Since our dataset is imbalanced and every class is equally important, we chose the Macro Averaged F1 score as the performance metric. According to the Classification report\ref{classification_report}, four out of five classes have achieved a F1 score above 95\%. The F1 score for the "product" class is 87\%. The reason for this low F1 score may due to the overfitting problem of the model. There are 322344 unique words that are labeled as "product", which is only 30\% of the overall "product" class size. The size of the unique data is too small, and there are not enough data samples to accurately represent all possible input data values. Therefore, the model is unable to generalize well on the novel data of the "product" class. This issue is a potential future work we will address in the last section.

Among the 64982 CVE sentences in the test dataset, the Bert model predicts 95\% of the CPE entities correctly on average. We have manually checked the dataset with wrong prediction result. There are two main causes for the errors.
\begin{enumerate}
	\item The original label is wrong.
	As shown in the Figure \ref{error_analysis_ground_truth_wrong}, the sentence is "Integer overflow in . . . SP3 allows remote attackers to execute arbitrary queries via a crafted . file that triggers an overflow allocation - size error , aka " . Integer Overflow Vulnerability .". The model predict the "." at the fourth position of the sentence as "O" entity. However, the Ground Truth label of the "." word in the entity is "B-vendor". The model predicts the correct result, while the original dataset is wrong.
	\begin{figure}[]
		\centering
		\includegraphics[width=7cm]{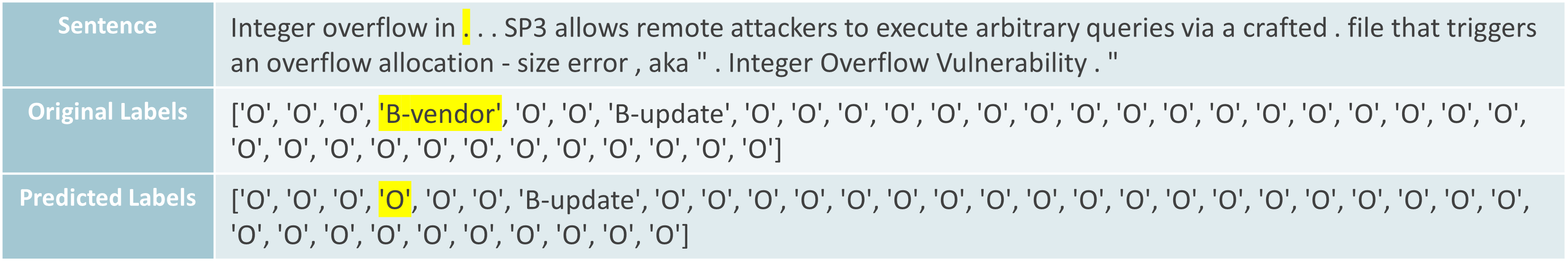}
		\caption{Ground Truth is Wrong}
		\label{error_analysis_ground_truth_wrong}
	\end{figure}
	\item The dataset treat the two continuous symbols as one single token, while the model predicts the two continuous symbols as two separate tokens.
	As shown in the Figure \ref{error_analysis_one_two_tokens}, the sentence is "REST ING endpoints j Jenkins 2 .. 218 and earlier , Java 2 .. 204 .. 1 and earlier were prone to clickjacking attacks ..". The model treats the two continuous symbols ".." as two separate tokens and gives two separate labels -- "O" and "O". However, the dataset treat the two continuous symbols as one single token, thus corresponds to one single label "O". This problem can be solved with better quality of training dataset.
	\begin{figure}[]
		\centering
		\includegraphics[width=7cm]{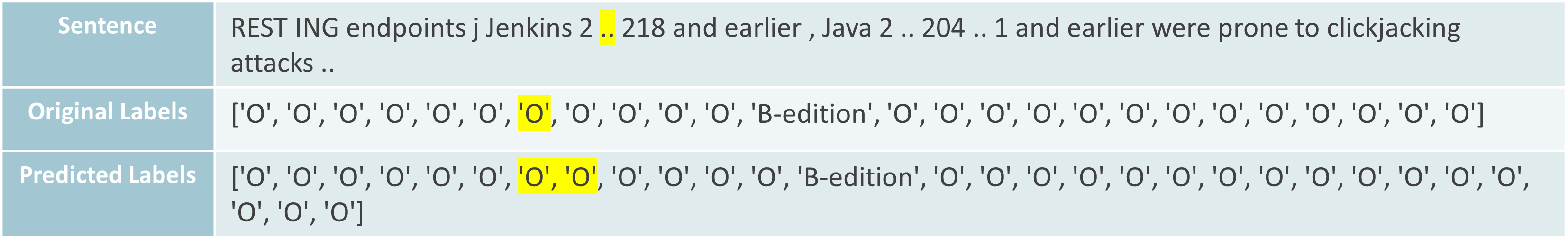}
		\caption{Different number of tokens}
		\label{error_analysis_one_two_tokens}
	\end{figure}
\end{enumerate}

\section{Conclusion \& Future Work}
\label{sec:Conclusion Future Work}
This paper designs an automated CPE labeling system called CPE-Identifier. The system can help the cybersecurity analysts automatically annotate new CVE summaries and identify key CPE entities, such as edition, product name, vendor name, version number, and update number. We have demonstrated the automated data annotation and augmentation processes by building two deep learning models. We have also compared the predictive performance of the three SoTA NER models when dealing with CVE summary text. We have concluded that the best model is the Bert model, which achieves an F1 score of 95.48\% and outperforms the baseline model from past related work. We have also built a Graphical User Interface that allows the cybersecurity analysts to use the CPE-Identifier system conveniently and zoom into the respective CPE entities accurately and efficiently. This paves the way for further discussions on various security operations to accelerate the incident handling processes when a new vulnerability emerges.

Regarding the limitations of the proposed approach, it is noteworthy that while the performance metrics of the Data Annotation model are commendable, the automated labeling process applied to the Named Entity Recognition (NER) dataset might introduces a degree of noise into the resultant dataset. This, in turn, exerts an adverse impact on the training process of the XLNet, BERT, and GPT-2 NER models. Therefore, for future research endeavors, we advocate for a meticulous manual examination of the labeled dataset prior to its utilization in training the aforementioned NER models. Additionally, we recommend enhancing the Data Annotation model by training it on more recent datasets, as it is currently fine-tuned on data spanning from 2010 to 2013. This limited temporal scope may result in suboptimal performance when annotating contemporary data, where novel technical terminologies frequently emerge.

Furthermore, in light of the increasing prominence of Large Language Models (LLMs) such as the GPT series, LlaMa LLMs, and the PaLM LLMs, we propose collaboration with these advanced language models in our future research endeavors to harness their capabilities effectively.

\bibliographystyle{apalike}

\end{document}